\documentclass[aps,prb,twocolumn,superscriptaddress]{revtex4-1}

\usepackage{graphicx}
\usepackage{amsmath}
\usepackage{bbold}
\usepackage[T1]{fontenc}

\begin{document}

\title{Water Radiolysis by Low-Energy Carbon Projectiles 
from First-Principles Molecular Dynamics}

\author{Jorge Kohanoff}
\affiliation{Atomistic Simulation Centre, Queen's University Belfast, 
                Belfast BT7 1NN, Northern Ireland, UK.}
\affiliation{Department of Earth Sciences, University of Cambridge, 
                Cambridge CB2 3EQ, UK.}

\author{Emilio Artacho}
\affiliation{Department of Earth Sciences, University of Cambridge, 
                Cambridge CB2 3EQ, UK.}
\affiliation{Theory of Condensed Matter, Cavendish Laboratory, 
        University of Cambridge, Cambridge CB3 0HE, UK.}
\affiliation{CIC Nanogune and DIPC, Tolosa Hiribidea 76, 20018 San 
         Sebasti\'an, Spain.}
\affiliation{Basque Foundation for Science Ikerbasque, 48013 Bilbao, Spain.}


\date{17 February 2017}

\begin{abstract}

 Water radiolysis by low-energy carbon projectiles is studied by
first-principles molecular dynamics.
  Carbon projectiles of kinetic energies between 175 eV and 2.8 keV
are shot across liquid water.
  Apart from translational, rotational and vibrational excitation,
they produce water dissociation.
  The most abundant products are H and OH fragments.
  We find that the maximum spatial production of
radiolysis products, not only occurs at low velocities, but also well
below the maximum of energy deposition, reaching one H every 5~\AA\ at
the lowest speed studied (1 Bohr/fs), dissociative collisions being
more significant at low velocity while the amount of energy required
to dissociate water is constant and much smaller than the projectile's
energy.
  A substantial fraction of the energy transferred to fragments,
especially for high velocity projectiles, is in the form of kinetic
energy, such fragments becoming secondary projectiles themselves.
  High velocity projectiles give rise to well-defined binary
collisions, which should be amenable to binary approximations.
  This is not the case for lower velocities, where
multiple collision events are observed.
  H secondary projectiles tend to move as radicals at high velocity, 
as cations when slower.
  We observe the generation of new species such as hydrogen peroxide
and formic acid.
  The former occurs when an O radical created in the collision process
attacks a water molecule at the O site.
  The latter when the C projectile is completely stopped and reacts
with two water molecules.
\end{abstract}


\maketitle

\section{Introduction}

  Water dissociation and the formation of other molecules by the action of 
radiation is one of the most important radiolytic processes, and has been 
studied for over a century by many authors.\cite{sonntag_87} While the main
interest in the subject is traditionally related to biological implications,
\cite{sonntag_87,hunniford_2007} and to nuclear reactor design,\cite{nuclear_08} 
it recently came into focus also within the energy context, due to the possibility of 
generating hydrogen at low cost.\cite{grimes_08} 
We will focus this study on ionic 
projectiles, and will not consider electromagnetic radiation. The two main 
natural occurrences of ions are: in space in the form of cosmic rays (mostly protons, 
$\alpha$-particles and electrons), and as products of radioactive decay in 
radionuclides. However, high-energy ions can also be produced in accelerators 
and used as radio-therapeutic tools (hadron-therapy). In either case, 
it is of major interest to understand, at the microscopic level, how do protons, 
$\alpha$-particles and heavier ions like C$^{+q}$ interact and split water or,
in a biological context,  produce reactive fragments that induce biological 
end-point effects such as DNA damage.

Most of these particles are very energetic (keV to MeV). When 
water is exposed to radiation of this nature the main effect is ionization, 
whereby electrons in the water orbitals are removed. The result is a 
characteristic distribution of secondary electrons whose kinetic energy peaks at low 
kinetic energies and then decreases monotonically.\cite{wilson_84,emfietzoglou_07} 
Other collision channels such as 
ion-molecule direct impact have exceedingly small cross sections in this
regime, and can be ignored. The ionization regime can be described 
quite well in terms of binary collisions with individual water molecules 
(gas phase) where the electronic structure is corrected for the influence of 
the environment (condensed phase).\cite{gaigeot_07} The information on 
scattering cross sections can then be used to study radiation tracks
via Monte Carlo simulations.\cite{gaigeot_07,cobut_98}

As ions travel through the medium ionizing the water, they gradually lose their
energy. 
Initially, the ionization cross section is small, but when their velocity 
approaches that of the electrons in the water orbitals, a resonance phenomenon 
takes place and a peak in the absorbed dose is observed (Bragg peak), which 
for carbon corresponds to a state of charge approximately 
C$^{3+}$.\cite{solovyov_09} Beyond the Bragg peak, the ionization 
cross section and the velocity of the ions rapidly decrease while the ions
capture additional electrons. 
Below a certain threshold, ionic 
projectiles do not have enough kinetic energy to ionize water. The electronic
excitation channel remains open, but only briefly. Water is an electronic 
wide gap insulator like LiF, for which the existence of a projectile-velocity 
threshold for electronic excitation has been shown\cite{Bauer_09} (and partly 
understood.\cite{pruneda_07,artacho_07}) to be 
between 0.1 and 0.2 atomic units of velocity. 
  For carbon projectiles and using what learned for LiF, the electronic 
excitation channel should essentially close at energies of the order of 
4 keV. 
  Below this, the collision process should be predominantly adiabatic, 
meaning that electrons remain in the instantaneous ground state corresponding 
to the nuclear configuration. 

This regime is extremely interesting for various reasons. In comparison with 
the ionization regime, it has always been assumed that low-energy ions produce 
comparatively little (if any) damage but, in fact, it remains largely unexplored. 
Within the radio-therapeutical context, where the energy of the incoming ions 
is adjusted to optimize the energy deposition, this region occurs beyond the Bragg 
peak. 
In the adiabatic regime, energy is transferred directly into 
translational, rotational and vibrational degrees of freedom of the target 
molecules. If a sufficient amount of energy is deposited into vibrations, 
then water molecules can dissociate thus originating 
various fragments like OH$^-$, H$^+$ and O$^-$ ions, and 
OH$^\bullet$, H$^\bullet$ and O$^\bullet$ radicals. These, in turn, are 
reactive species that tend to either recombine, if the fragments have not 
travelled too far away from each other, or to associate with other fragments 
and with water molecules, thus giving rise to more complex entities such as 
hydrogen peroxide (H$_2$O$_2$) and HO$_2^\bullet$. Within this regime, the 
binary collision hypothesis becomes questionable, and the use of gas phase 
calculated or measured scattering cross sections requires careful validation.

In this work we study the collision and chemical processes that occur when
a carbon projectile traverses a slab of liquid water, by first-principles 
molecular dynamics simulations, within the adiabatic regime. 
We show that the largest density of damage occurs at 
projectile energies well below the peak in energy deposition. This is consistent
with experimental findings of low-velocity C$^+$ ions impinging on DNA plasmid
samples.\cite{hunniford_2007} The explanation
is that at higher velocities collisions are more infrequent, but most of the 
transferred energy goes into kinetic energy of the collision products.
Therefore, the amount of energy stored into locally reactive species is
approximately constant, but at lower velocities it is spatially more densely
distributed.

\section{Methods}

The work is based on first-principles molecular-dynamics simulations (FPMD).
These have been carried out with the Siesta code,\cite{siesta_02} which is based on 
density functional theory and pseudopotentials, and uses numerical pseudoatomic
orbitals as a basis set. A reasonable description of hydrogen bonding in water
\cite{marivi_04} is given by the generalized gradient approximation as proposed in
\cite{pbe_96} (GGA-PBE). The pseudopotentials and basis functions have been validated 
previously in a study of bulk liquid water.\cite{marivi_04}
A previous study on the behaviour of pseudopotentials for binary 
collisions\cite{Pruneda_04} supports the pseudopotentials used here
for the scale of energies and distances attained.

  A bulk sample of 128 water molecules was equilibrated with FPMD at a
temperature of 300 K over 10 ps in a cubic box of side 15.736 \AA, under 
periodic boundary conditions. 
  For practical purposes a water slab was generated, which allows for a better 
definition of the initial and final states of the sample and the projectile, as well as of the
energy loss of the latter.
  It should be remarked, however, that it is the bulk behaviour what is explored
here, not the thin-film or surface effects.
  For that purpose, the bulk liquid box was expanded to twice that length along 
the $x$-direction without modifying the atomic positions, the extra space 
playing the role of a vacuum slab.
  The water slab so created was further equilibrated with FPMD for 0.5 ps, which
is sufficient time for the molecules to accommodate in the short range, but 
intentionally insufficient for larger range relaxations (surface layering etc).
  Two snapshots were extracted from the equilibrated run, and the shooting of a carbon 
projectile through the water slab was simulated for a set of 36 different initial 
conditions for each snapshot and initial velocity, thus totalling 72 trajectories
per velocity. The projectile was shot 
from locations in vacuum distributed in a uniform square grid in the 
$yz$-plane, covering the whole surface area of the slab. The initial velocity of 
the projectile was chosen perpendicular to the slab, i.e. along the $x$-direction. 
The overall charge of the system was set to +1$e$, and was compensated with a uniform 
neutralizing background as required for extended periodic systems (C$^+$ is
the relevant charge state at low velocities \cite{hunniford_2007}). We studied four 
different initial velocities, 0.025, 0.05, 0.075 and 0.1 a.u. corresponding to
kinetic energies 175 eV, 700 eV, 1.6 keV and 2.8 keV, respectively. The latter 
is close, yet below the estimated limit of validity of the adiabatic hypothesis. The 
Newtonian equations of motion for the nuclei were integrated using a velocity 
Verlet algorithm with a time step of 0.1 fs.

\section{Results and Discussion} 

\begin{figure*}
  \includegraphics [width=0.3\textwidth]{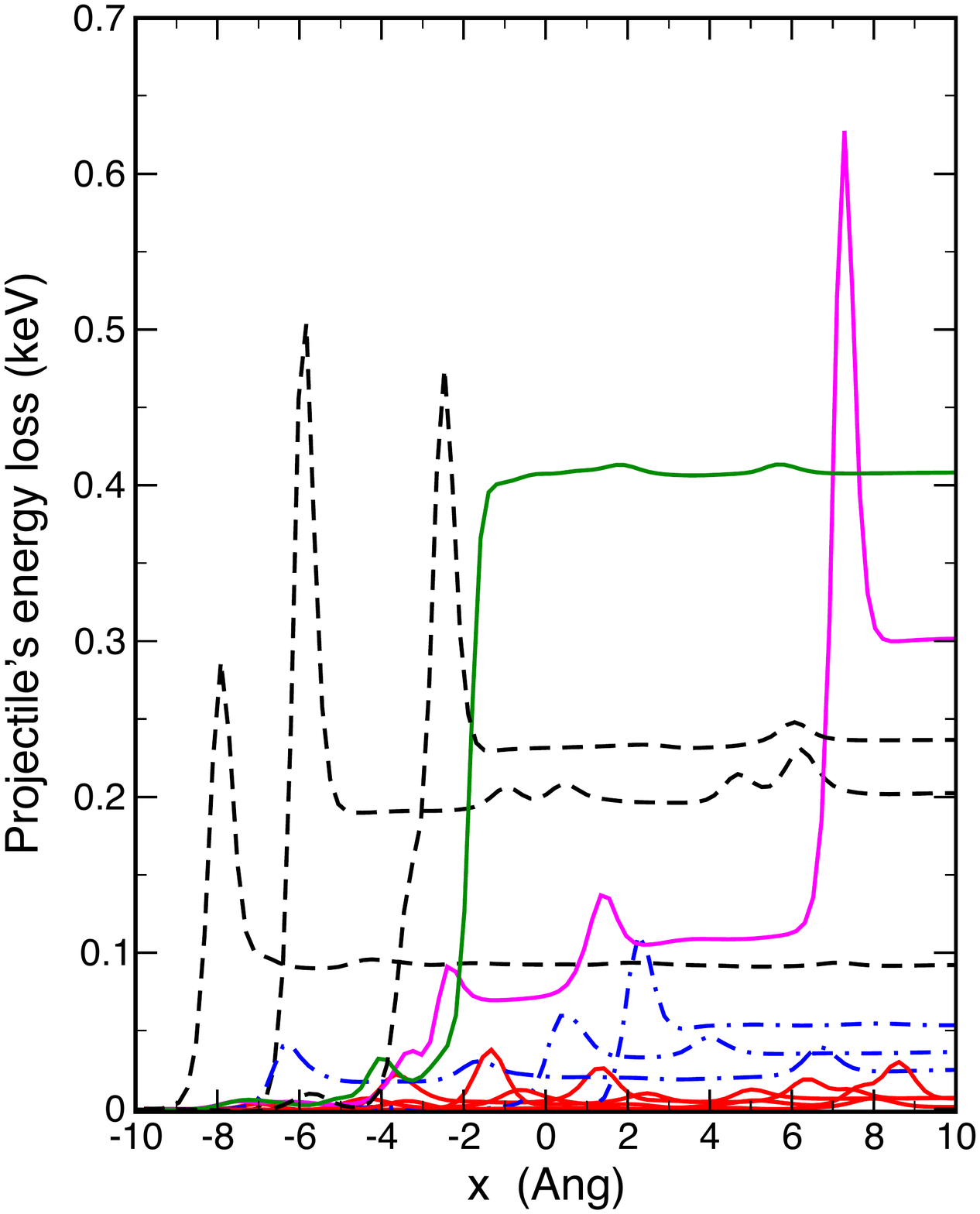}
  \includegraphics[width=0.3\textwidth]{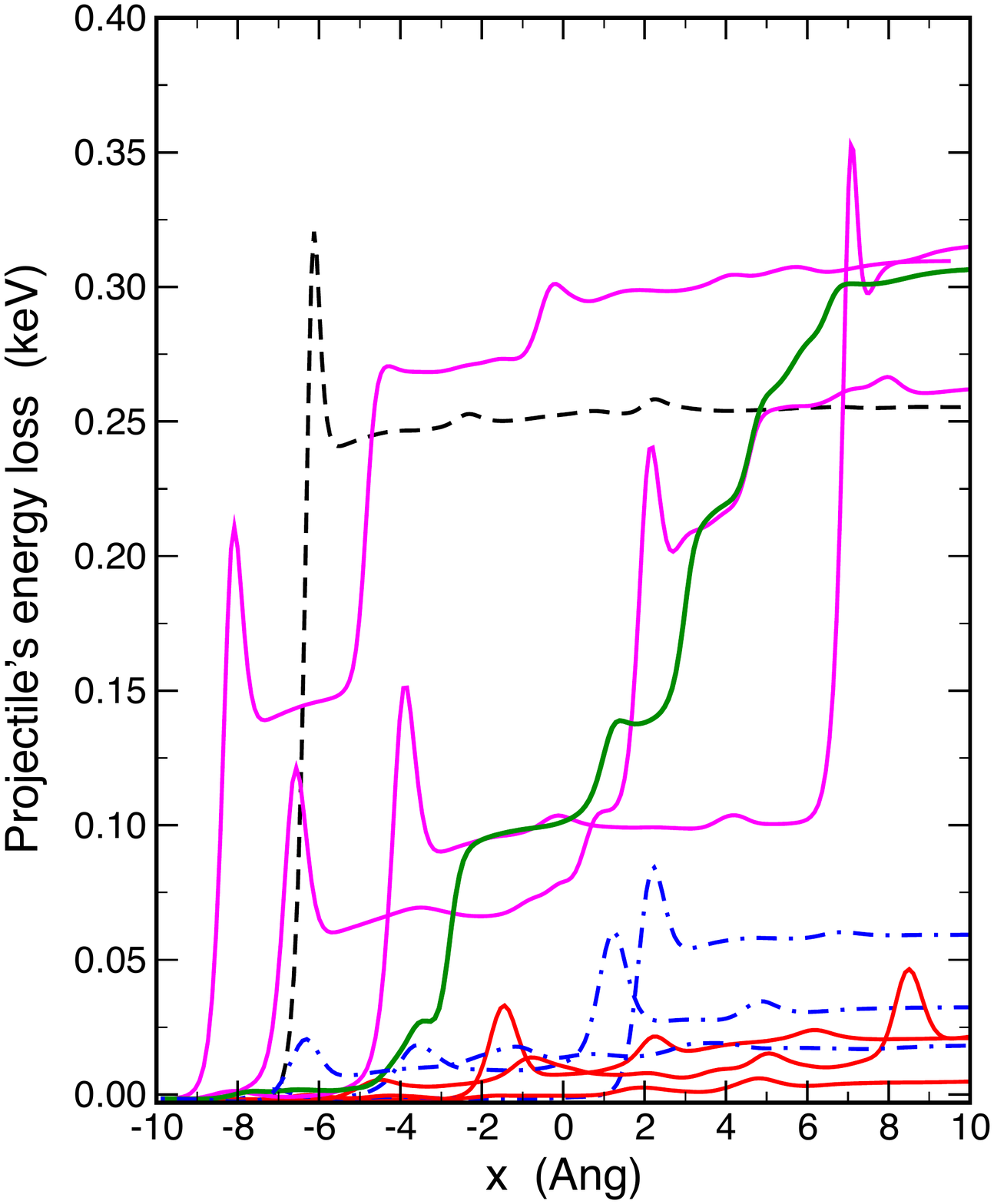}
  \includegraphics[width=0.30\textwidth]{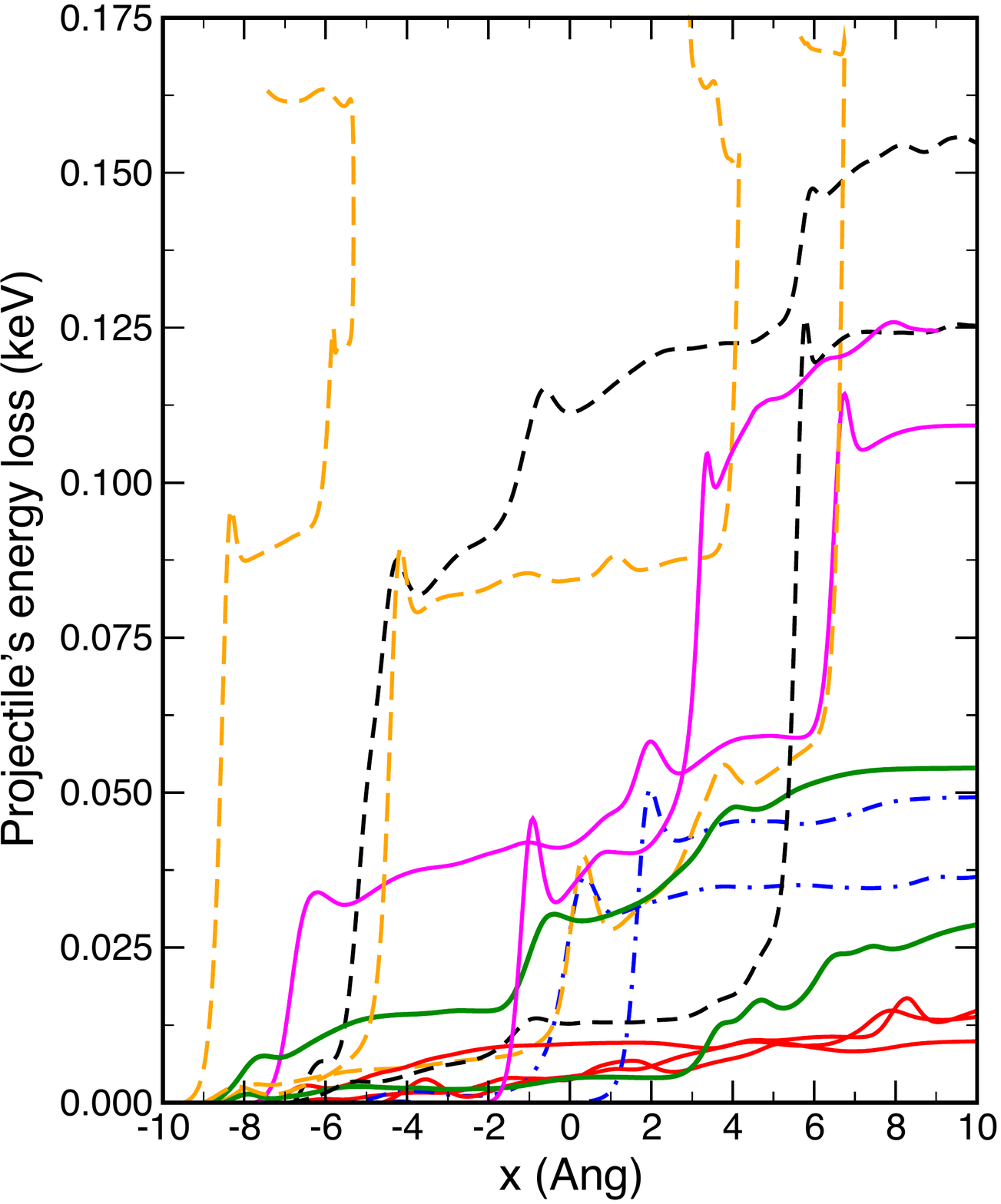}
  \caption{Energy lost by the projectile to the medium as a function of the 
           distance traversed in the water slab. Left panel: Initial energy 2.8 keV.
           Thin solid (red) lines represent non-dissociative trajectories. 
           Dashed (black) lines involve the complete dissociation of a 
           single molecule, and dot-dashed (blue) lines are for single H 
           dissociation from a water molecule. The two thick solid
           lines represent rare events of multiple, sequential dissociation 
           (magenta) and large kinetic energy transfer to a single H (green).
           The pronounced peaks occur when the projectile reaches very close
           to the target, but this energy is then partially restored.
           Middle panel: Initial energy is 0.7 keV. Lines and colors as in left
           panel, except for the thick solid (green) line, which corresponds
           to a trajectory where the projectile transfers energy continually to
           the medium. Such cases cannot be described in terms of binary collisions. 
           Right panel: Initial energy is 0.175 keV. Lines and colors as middle panel.
           Long-dashed orange lines represent trajectories where the projectile
           is stopped completely within the water slab. The peaks just before each
           collision are due to the projectile experiencing the 
           repulsion of a target atom at close proximity.}
\label{trajectories}
\end{figure*}

We first analyze the various types of trajectories and collision events by plotting
the evolution of the energy loss of the projectile to the water for some selected, 
representative simulations. Fig \ref{trajectories} shows that for an initial kinetic 
energy of 2.8 keV, collisions with the water molecules are not very 
frequent, thus suggesting that the cross section is relatively small. 
Nevertheless, for small impact parameters, collisions 
are quite dramatic, invariably leading to the dissociation of the water molecule, either 
singly (C+H$_2$O$\to$C+OH+H) or doubly (C+H$_2$O$\to$C+O+2H). The energy loss 
to the fragments is quite significant, between 25 and 50 eV for single dissociation 
events (dot-dashed lines in Fig \ref{trajectories}), and 100 to 150 eV for complete 
dissociation (dashed lines). Within our GGA-PBE and also experimentally,\cite{Blanksby_2003} 
the dissociation energy of the first H-atom in the water molecule is about 5 eV. 
The second H dissociates for an extra 4.4 eV.\cite{Blanksby_2003} These energies are 
always small compared to the total 
energy transferred, amounting to a 5-10\% at the most. The remainder is transferred as 
kinetic energy of the fragments, which are thus produced as hyperthermal 
species.\cite{deng_06} There are also many trajectories where the projectile traverses 
the sample without impacting any water molecule, and losing only a small fraction of 
its energy into non-dissociative channels such as vibrational excitation. An important 
observation is that, at such large kinetic energies, the projectile proceeds via a 
succession of individual collision steps, thus justifying the binary collision model.
Such individual collisions are evident in the sharp peaks exhibited by the curves
in Fig 1. There, the projectile experiences a collision with one of the atoms in the target.
This collision  has an inelastic and an elastic component, the latter being responsible for the 
peak. Kinetic energy is temporarily converted into potential energy, and then quickly 
recovered.

Simulations with initial energy of 1.6 keV are not qualitatively very different 
from those at 2.8 keV. Therefore, we next analyze trajectories generated for an 
initial kinetic energy of 700 eV (middle panel in Fig \ref{trajectories}). 
Now multiple collision trajectories are much more frequent.
The energy transfers are slightly, but not significantly smaller, i.e. around 100 
eV for dissociative events. This is a larger fraction of the initial energy, 
but not large enough to stop completely the projectile. There is a noticeable increase 
in the amount of secondary particles generated by dissociation, due to the enhanced 
cross section at lower velocities. By inspecting the trajectories
it can be observed that in this regime the projectile has time to {\it drag} the
particles, thus giving rise to less dramatic dissociation events where the
O-H stretching vibrational mode is excited beyond the point of no return
(see S1-S6 Videos in the Supporting Information).
In some cases this proton joins another molecule forming H$_3$O$^+$
while the remaining OH$^-$ reacts with a water molecule to form H$_3$O${_2}{^-}$.
One of the trajectories depicted in dot-dashed (blue) lines corresponds precisely
to the creation of this species, at an energetic cost of 16 eV for the projectile.

An even more interesting feature emerging at this velocity is the appearance of
trajectories where the energy loss is not by steps, but rather in an almost
continuous way by dissociating water molecules and exciting vibrational modes.
This is represented by the thick (green) solid line. This class of 
trajectories cannot be satisfactorily described in terms of binary 
collisions, thus signalling the breakdown of binary collision models.

In the right panel we show the evolution of the projectile's kinetic energy for
an even lower initial energy of 175 eV. Now the majority of the trajectories 
presents a continuous energy loss to vibrations and rotations which, due to the 
scale of the energy transfers, becomes more evident. A new type of trajectories
appears at this velocity, where the 
projectile is completely stopped and thermalized in the medium (long-dashed orange 
lines). By inspecting these trajectories, it turns out that the energy transfer 
in dissociative collision events is still quite large, up to 100 eV (see dashed 
black and orange lines). Therefore, after two or three such events (orange lines), 
the projectile has exhausted all its kinetic energy and stops. 

As before, the energy to 
dissociate the water molecules is a small fraction, and most of the kinetic
energy is transferred to the hyperthermal fragments. In fact, in some trajectories 
the C is stopped, transferring its energy to an O atom or an OH group, which then
assumes the role of the projectile. Other types of events observed are the abstraction
of a H-atom by the projectile to form a travelling C-H group
(see S1-S6 Videos in the Supporting Information).
  Here we restrict ourselves to an inventory of a variety of events that 
have been observed in the present simulations. 
  We do not attempt to estimate the probability of occurrence of such 
events, which would require the careful evaluation of free energies and 
integration over impact parameter and water orientations, apart from 
statistical averaging.

\begin{figure}
\begin{minipage}{0.5\textwidth}
  \includegraphics[height=.27\textheight]{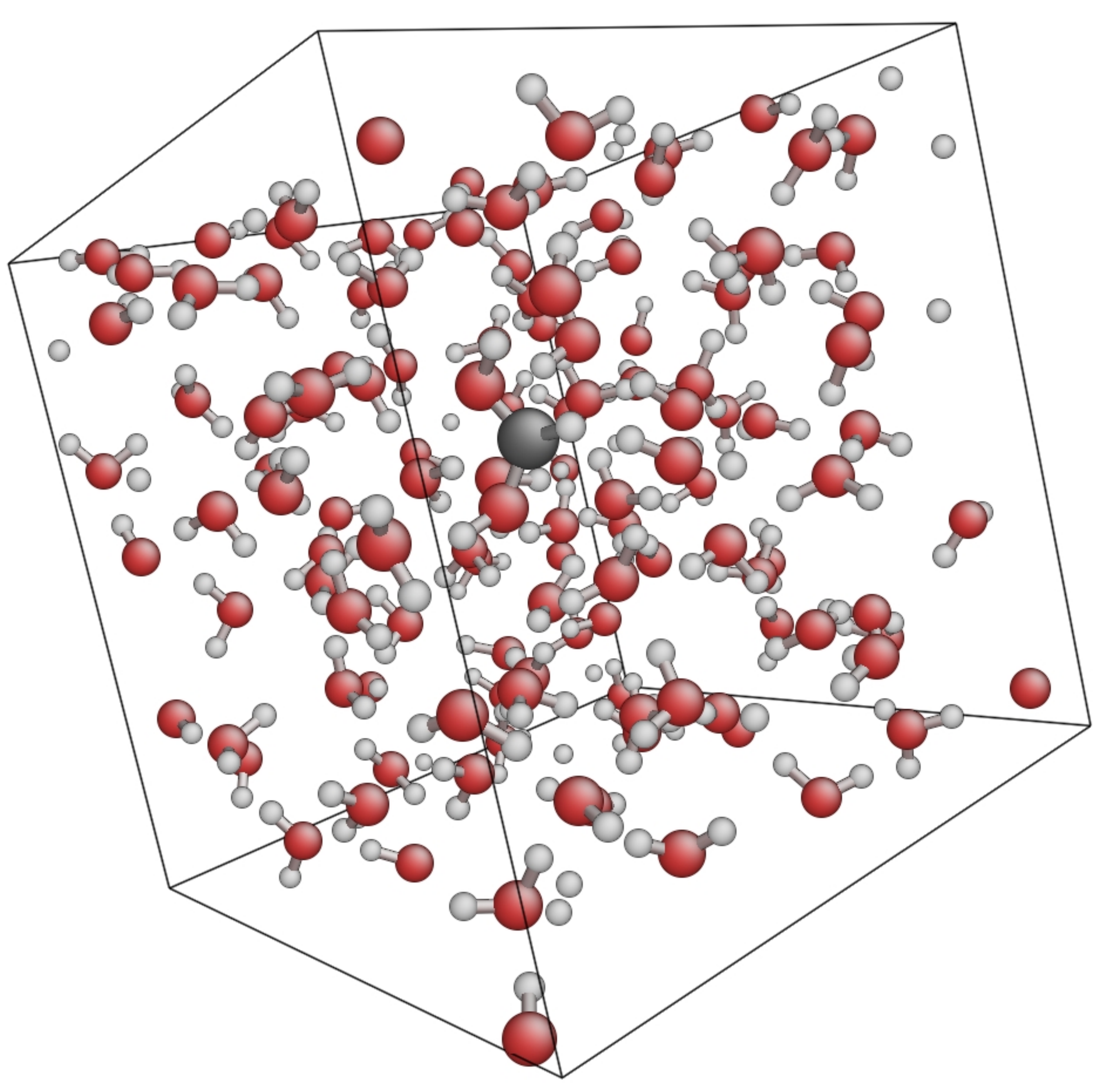}
\end{minipage}
\begin{minipage}{0.5\textwidth}
  \begin{center}
    \includegraphics[height=.22\textheight]{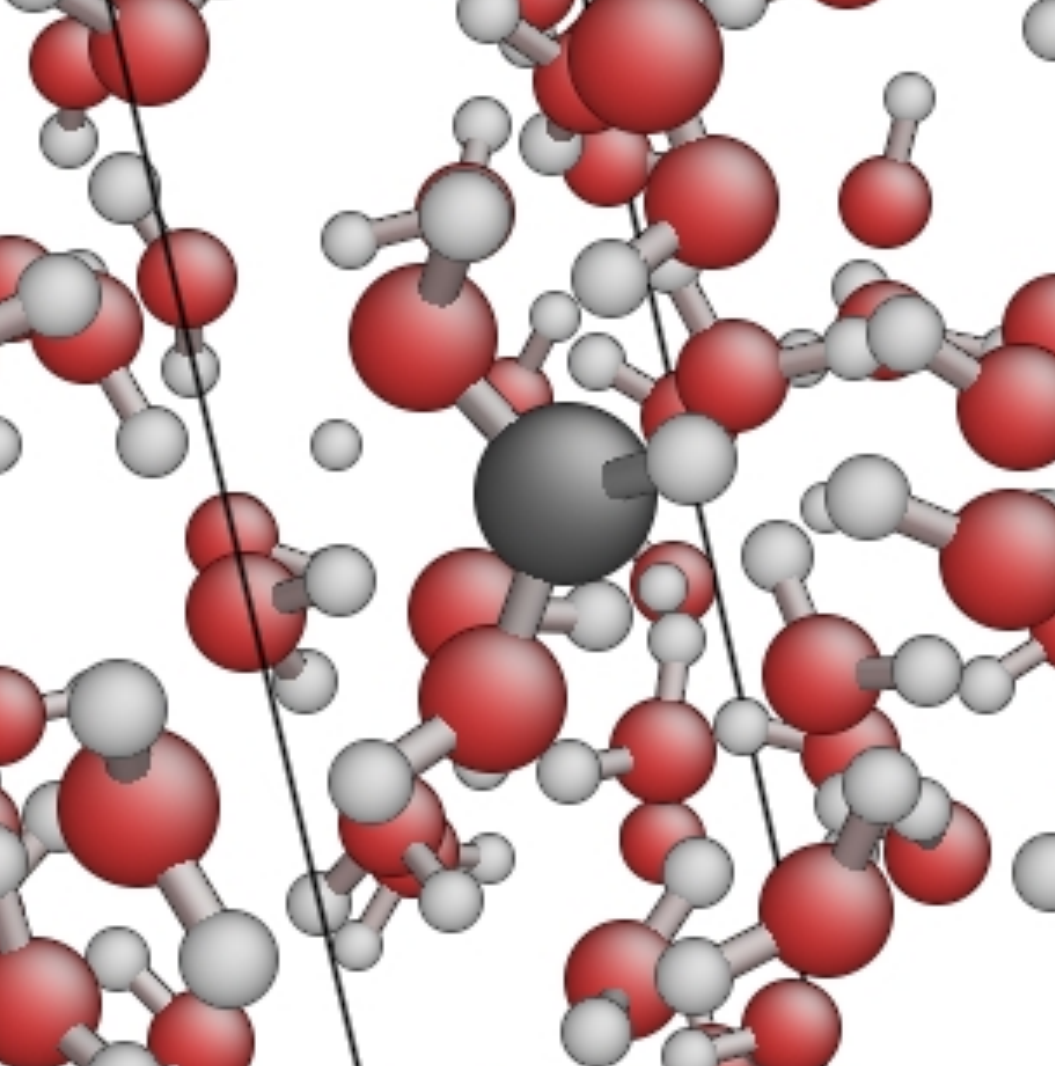}
  \end{center}
\end{minipage}
  \caption{The formation of an organic chemical species by reacting carbon with
           water in the condensed phase. In this case the species is dihydroxymethyl 
           radical (closely related to formic acid), which was formed during a 
           first-principles MD simulation.}
\label{formic}
\end{figure}

It is important to remark that, since the dissociation energy is small compared 
to the energy transfer, the generation of secondary particles is not hindered at
low energies. On the contrary, it appears enhanced with respect to
the previous cases. The difference is that the H, O or OH fragments have lower
kinetic energies, and are thus more prone to stop and react with other water
molecules or fragments. In fact, in some cases we have observed the formation
of new chemical species such as hydrogen peroxide H$_2$O$_2$, which is known
to be one of the most common products in irradiated water within the ionization
regime.\cite{sonntag_87} Interestingly, this phenomenon occurs here via the
attack of a water molecule by an oxygen radical arising from a fully
dissociating collision, unlike the conventionally accepted mechanism of
the reaction between two OH radicals.\cite{sonntag_87}

In order to study the situation after a C projectile has stopped, we have conducted 
simulations of a C$^+$ ion in bulk water (128 molecules in a cubic box without vacuum). 
Here we have observed the formation of organic molecules such as the dihydroxymethyl 
radical HCO(OH)$_2^{\bullet}$, where the carbon impurity has reacted with two 
water molecules, as shown in Fig~2.
  The formation of this species has also been observed in a recent study of
irradiation of water clusters at 30 K, within the astrochemical context. \cite{emmet_2014}

To analyze the distribution of species generated by the projectile, 
we have monitored the coordination number of the O and 
H-atoms along the track (see Fig~3) by considering only 
atoms within a specified distance cutoff, 
which we set to 1.2 \AA, except for H-H pairs which we set to 0.8 \AA. Below, $Z_X(Y)$ 
indicates the number of $Y$-atoms coordinated to an $X$-atom.Thus, $Z_O(Y)=0$ 
for all $Y$ implies an isolated oxygen atom or ion (lower left panel of Fig~3), 
$Z_O(H)=1$ indicates a hydroxyl anion or radical 
if $Z_O(O)=0$, or hydrogen peroxide if $Z_O(O)=1$ (lower right panel). If $Z_O(H)=2$ we 
have the usual water molecule (not shown), while $Z_O(H)=3$ represent hydronium ions, 
H$_3$O$^+$ (upper left panel). In the case of H-atoms, $Z_H(Y)=0$ for all $Y$ implies 
hydrogen atoms 
or protons (upper right panel), while $Z_H(H)=1$ is for H$_2$ molecules (not shown). 
$Z_H(O)=1$ can correspond to hydroxyl or hydronium groups, and $Z_H(O)=2$ suggests 
H$_5$O$_2^+$ groups (not shown). The quantities reported are the result of averaging 
over the 72 trajectories.
  The concentration of water fragments obtained is very substantially larger than
the one observed at room conditions ($\sim 10^{-7}$M at neutral pH).
  Water splitting processes (mostly into H$^+$ and OH$^-$) are also induced by 
high temperatures and pressures\cite{Holzapfel1969,Cavazzoni1999},
although a substantial splitting ($\sim 1$ mol \%) is not observed below 
2000 K and 12 GPa.

\begin{figure}
\begin{center}
  \includegraphics[width=.5\textwidth]{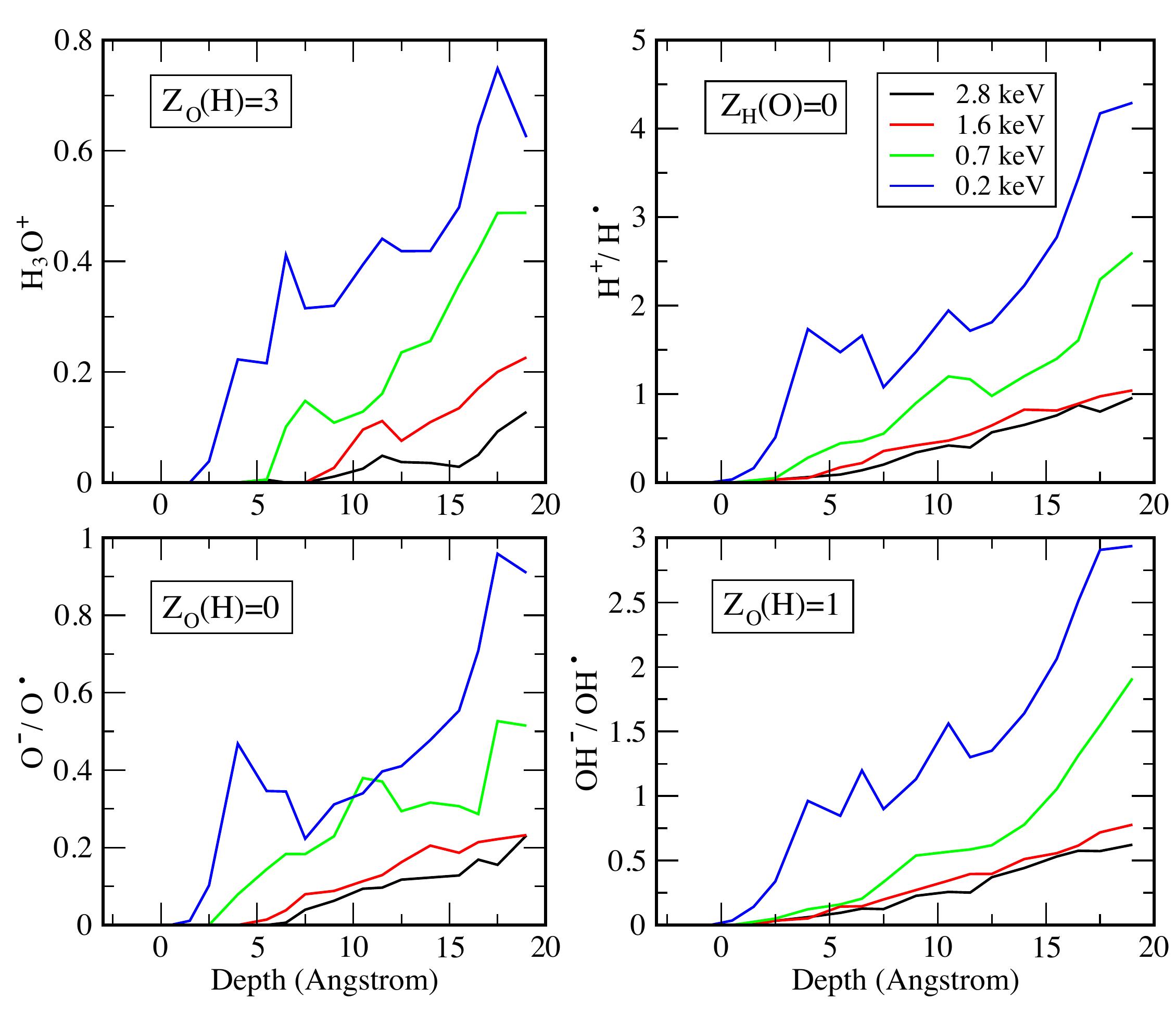}
\end{center}
\caption{Evolution of the average number of secondary species 
generated during the passage of the projectile through the water 
slab, as a function of the spatial coordinate of the projectile.
  Upper left panel shows O-atoms coordinated to 3 H, lower left 
is for isolated O-atoms, upper right for isolated H-atoms, and 
lower right for O-atoms coordinated to 1 H-atom. 
  The latter represents OH groups and also the less frequent 
H$_2$O$_2$ species.
  Colors correspond to the various initial conditions: 2.8 keV 
(black), 1.6 keV (red), 0.7 keV (green), and 0.17 keV (blue).} 
\label{fragments}
\end{figure}

Interestingly, the density of secondary fragments generated along the track 
{\it increases} with decreasing velocity 
of the projectile. This is due to a resonant behavior of low-energy projectiles 
with molecular vibrations and rotations. At high velocities vibrations do not have the time 
to react to the passage of the projectile. Collisions are effective only for 
small impact parameters, being dissociation a secondary effect of the energy
transfer. The spatial rate of fragment generation 
increases for slower projectiles up to a point in which dissociation events become 
less viable due to the small kinetic energies involved. The two most abundant
fragments are H and OH groups. At the lowest energy studied here, the former
are generated at a rate of one every 5~\AA, while for OH it is one every 7~\AA, the
difference being due to the double dissociation events.
This means that the projectile dissociates basically every molecule that finds in 
its way. This rate decreases with increasing energy, down to one H every 20~\AA~ 
and one OH every 30~\AA~at 2.8 keV. 

Hyperthermal fragments become secondary projectiles 
that originate secondary collision cascades, rather than thermalizing 
with the medium and locally heating the water.
We have analyzed some of these trajectories. 
Interestingly, while the OH and O fragments 
rapidly experience other collisions, the mean free path for the H-atoms is much 
longer and they travel almost freely through the
water film.
Hence, much larger simulation boxes would be required to study radiolysis by 
protons or $\alpha$-particles.

A very important quantity is the charge state of the projectile and the secondary
fragments. It determines the strength of the interaction with the medium, and
as such it is a crucial ingredient to compute cross sections in binary collisions. 
In addition, the charge state determines the reactivity of low energy fragments,
and thus their damaging power, e.g. it is not the same an OH$^\bullet$ radical
than a OH$^-$ hydroxyl anion, or a proton and a hydrogen atom. In particular,
neutrals tend to diffuse faster and longer than charged species. 

The coordination numbers presented above do not carry information 
about the charge state. 
Therefore, we have to resort to some other method. 
Although the charge state 
of an atom or molecule in the condensed phase is not a well-defined quantity,
Mulliken populations are a useful quantity to look at. 
The ambiguities intrinsic to any population (and to Mulliken's) are small when 
calculating the population of a whole molecule, radical or ion. 
We have done this especially 
for the products of dissociative collision events, to understand 
whether the H and O atoms leave as neutral or charged species.

It is important to remark that all the processes considered here are adiabatic, 
in the sense that the charge is the result of the self-consistent determination 
of the charge density at each nuclear configuration. A hydrogen atom in a water 
molecule will exhibit a Mulliken charge of about 0.7, while a neutral hydrogen
will carry a charge of 1 and a proton a zero charge. Since in the condensed phase 
a hydrogen atom can use the basis functions of other neighboring atoms, the charge 
will be generally smaller than 1. 
We observe that, when hyperthermal H
or O atoms leave the slab, they do so as neutral atoms. However, when they 
remain in the slab, we observe the trend that 
hyperthermal hydrogens generally move as neutral atoms (with populations 
between 0.9 and 1 electron), while slow ones tend to move as protons (with 
populations smaller than 0.4 electrons, which account for the contribution of 
the H basis orbitals to the description of the electrons in close-by molecules). 

72 trajectories per velocity value may not be sufficient to obtain good statistics
for some properties. 
Nevertheless, there are some averaged quantities that are meaningful as 
qualitative indicators of the general behavior of the system. We first show the 
distribution of final kinetic energies of the projectile (inset to Fig~4). 
This has been calculated by monitoring its velocity once it has passed 
through the water slab and no more collision events occur, and collecting them in a 
histogram. Consistently with the previous observation that the magnitude of the 
energy transfer does not depend strongly on the energy of the projectile, the 
width of the distribution is practically energy-independent, except for the 
lowest energy where some projectiles are completely stopped. 
This information can be used to estimate the nuclear stopping power as the 
ratio between the energy deposited along the trajectory and the width of the 
slab. This is an estimate of a differential quantity averaged over a finite-width slab, 
which carries additional statistical uncertainties. Although it is not particularly 
accurate, it is qualitatively useful and we show it in Fig~4. 
It is tempting to conclude that there is a maximum in stopping power for energies 
around 2-3 keV, but statistical errors preclude a robust interpretation.

\begin{figure}
\begin{center}
  \includegraphics[width=.4\textwidth]{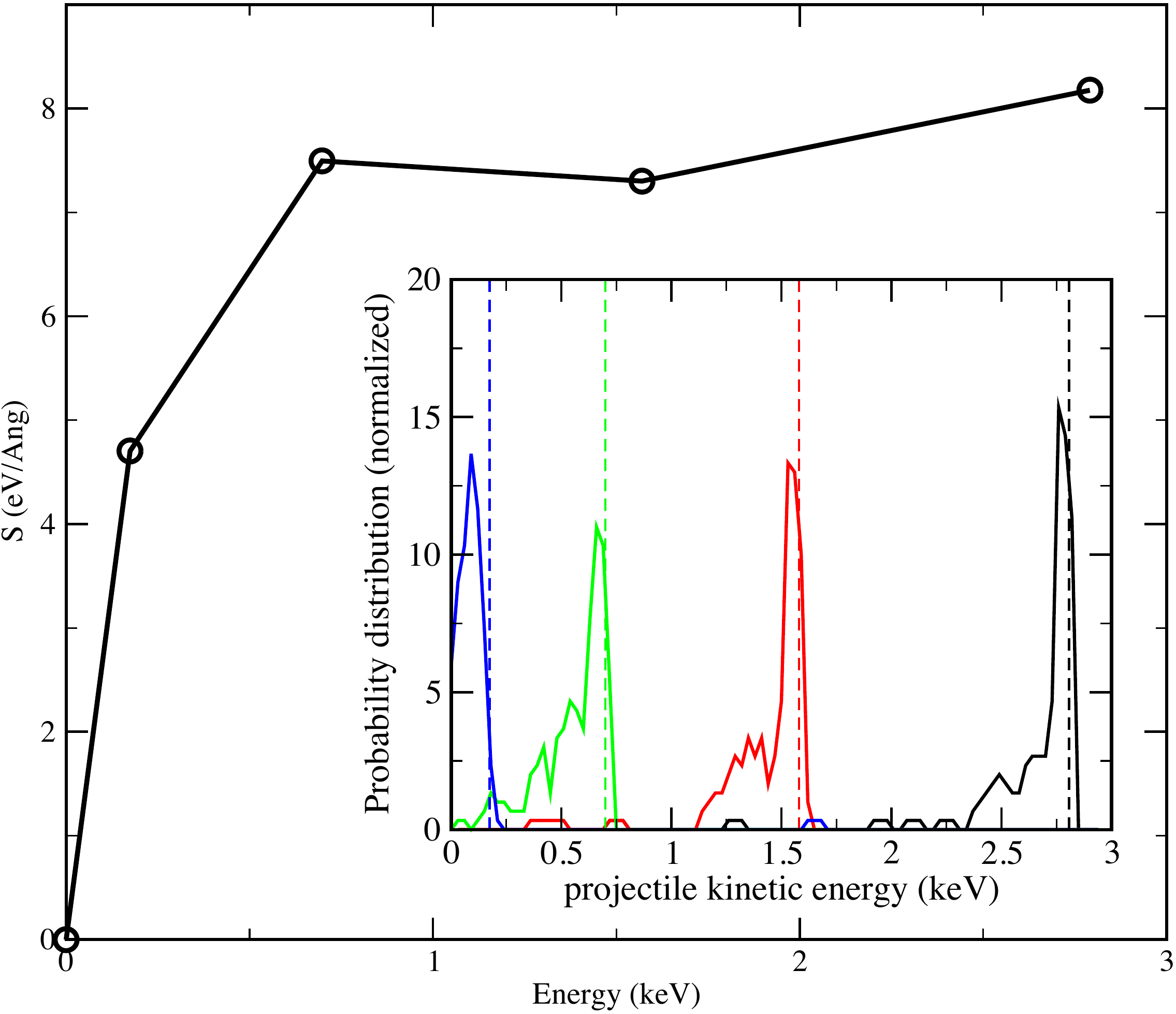}
\end{center}
\caption{Nuclear stopping power as a function of projectile kinetic energy.
  The inset shows the distribution of final kinetic energies of the projectile 
for the various initial conditions: $v_0=0.1$ a.u. (black), 
$v_0=0.075$ a.u. (red), $v_0=0.05$ a.u. (green), and $v_0=0.025$ 
a.u. (blue).
  Vertical dashed lines indicate the initial kinetic energy.}
\label{kinene}
\end{figure}

The rationale for such a maximum is that, qualitatively, the stopping power is 
determined by the cross section for the interaction of the projectile with individual 
molecules and the energy transferred in those collisions. While the former decreases
with increasing energy, the latter increases but at a lower rate. Theoretical
expressions for the location of this maximum do exist,\cite{zbl_85} but they require
the knowledge of the charge state of both projectile and target, and are not 
directly applicable to molecular targets. A crude estimate using available
theories gives a value lower than the one observed here.\cite{debrecen_08} 
It should be emphasized that a more precise and statistically better defined 
nuclear stopping power would need substantial extra computation, which is
beyond the scope of this paper, and could be the basis of future work.
Similarly, further work would be interesting using higher levels of theory,
whenever affordable. In particular, the introduction of self-interaction 
corrections (with spin polarization) would allow a better description of
radical species, when exploring the processes in a finer energy scale,
in addition to the explicit consideration of possible non-adiabatic effects. 
The main conclusions of this work, however, given the scales involved,
are well supported by the level of theory employed.



\section{Conclusions}

To summarize, the first-principles molecular-dynamics simulations
presented for water radiolysis due to low-energy C bombardment,
show high yields of H an OH fragments that act as secondary
projectiles, especially for the higher velocities explored.
They are however more abundantly produced for lower
velocities, in multiple collisions that cannot be described in
a binary framework.
  H secondary projectiles tend to move as neutral atoms (radicals)
at the higher velocities observed, while it tends to move as a cation
when slower.
New species are seen to be produced, such as hydrogen 
peroxide and formic acid. 
The former occurs when an O radical created in the 
collision process attacks a water molecule. The latter when the C projectile 
is completely stopped and reacts with two water molecules.

\section{Supporting Information}

\subsection*{S1 Video}
\label{S1_Video}
{\bf C$^+$ ion impinging on water with a kinetic energy of 175 eV.}  
The different colors indicate different chemical species.
There are six video files associated to this paper, corresponding to three
different initial velocities of the C$^+$ cation. All of them from the same initial
conditions, except the magnitude of the velocity. They all correspond to initial
position 29 within the 72 trajectories per velocity.

\subsection*{S2 Video}
\label{S2_Video}
{\bf C$^+$ ion impinging on water with a kinetic energy of 175 eV.} 
The different colors indicate different coordination.

\subsection*{S3 Video}
\label{S3_Video}
{\bf C$^+$ ion impinging on water with a kinetic energy of 700 eV.} 
The different colors indicate different chemical species.

\subsection*{S4 Video}
\label{S4_Video}
{\bf C$^+$ ion impinging on water with a kinetic energy of 700 eV.} 
The different colors indicate different coordination.

\subsection*{S5 Video}
\label{S5_Video}
{\bf C$^+$ ion impinging on water with a kinetic energy of 2.8 keV.}
The different colors indicate different chemical species.

\subsection*{S6 Video}
\label{S6_Video}
{\bf C$^+$ ion impinging on water with a kinetic energy of 2.8 keV.}
The different colors indicate different coordination.

\begin{acknowledgments}
We thank Marivi Fern\'andez-Serra for the pseudopotentials, bases and
equilibrated water configurations, and Kostya Trachenko for useful discussions 
and help with visualization tools. JK thanks the Wellcome Trust for a flexible 
travel award to support this research, and the Department of Earth Sciences at 
University of Cambridge for hospitality. 
EA ackowledges financial support from the following grants: Electron-Stopping, 
from the European Commission under the Marie-Curie CIG, program;  FIS2012-37549-C05 
from the Spanish Ministry of Science; and Exp. 97/14 (Wet Nanoscopy) from 
the Programa Red Guipuzcoana de Ciencia, Tecnolog\'{\i}a e Innovaci\'on, 
Diputaci\'on Foral de Gipuzkoa.
Simulations were performed in 
the High Performance Computer at the University of Cambridge, and at the HECToR 
facility under the UKCP consortium EP/F037325/1.
\end{acknowledgments}

\end{document}